\documentclass[11pt,preprint]{aastex}
\begin{document}

\title{Cryogenic Tests of Volume-Phase Holographic Gratings: I. Results
at 200 $K$}

\author{Naoyuki Tamura, Graham J. Murray, Peter Luke, Colin Blackburn,
David J. Robertson, Nigel A. Dipper, Ray M. Sharples, \& Jeremy
R. Allington-Smith}
\affil{Department of Physics, University of Durham, South Road,
 Durham, DH1 3LE, UK}
\email{naoyuki.tamura@durham.ac.uk}

\begin{abstract}

We present results from cryogenic tests of a Volume-Phase Holographic
(VPH) grating at 200 $K$ measured at near-infrared wavelengths. The aims
of these tests were to see whether the diffraction efficiency and
angular dispersion of a VPH grating are significantly different at a low
temperature from those at a room temperature, and to see how many
cooling and heating cycles the grating can withstand. We have completed
5 cycles between room temperature and 200 $K$, and find that the
performance is nearly independent of temperature, at least over the
temperature range which we are investigating. In future, we will not
only try more cycles between these temperatures but also perform
measurements at a much lower temperature (e.g., $\sim 80~K$).

\end{abstract}

\keywords{Spectrographs, gratings, near-infrared}

\section{Introduction}

Volume-Phase Holographic (VPH) gratings potentially have many advantages
over classical surface-relief gratings (Barden, Arns, \& Colburn 1998;
see also Barden et al. 2000), and are planned to be used in a number of
forthcoming instruments (e.g., AA$\Omega$; Bridges et al. 2002). While
applications to optical spectrographs only are currently being
considered, VPH gratings will also be useful to near-infrared
spectrographs if the performance at low temperatures is satisfactory. In
particular, its diffraction efficiency and angular dispersion should be
confirmed. Contraction of dichromated gelatin with decreasing
temperature could cause variations in the line density and profile of
diffraction efficiency (the thickness of the gelatin layer is one of the
parameters defining diffraction efficiency). Since cooling and heating
cycles might cause some deterioration of a VPH grating and reduce its
life time, we also need to see whether these characteristics vary with
the successive cycles.

\begin{figure}
\begin{center}
 \includegraphics[height=10cm,angle=-90,keepaspectratio]{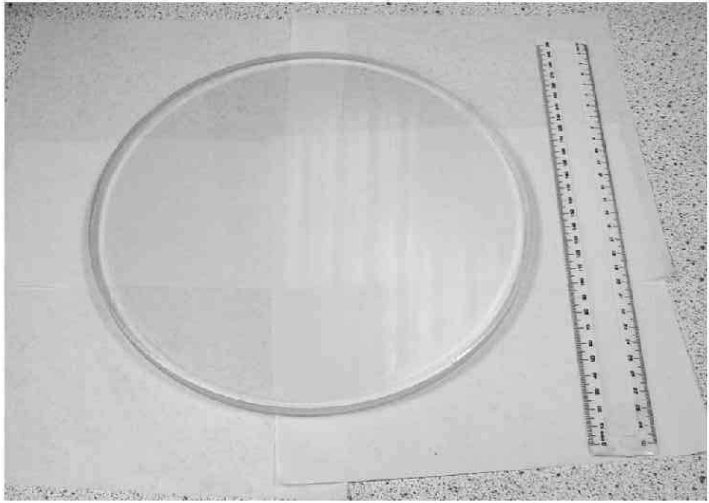}
 \caption{A picture of the sample VPH grating. The grating has a
 diameter of 250 mm with a line density of 385 lines/mm.}
 \label{vph}
\end{center}
\end{figure}

In this paper, results from measurements of a sample grating at 200 $K$
and at room temperature are presented. A picture of the grating
investigated is shown in Figure \ref{vph}. This grating was manufactured
by Ralcon Development Lab, and its diameter is about 25 cm. The line
density is 385 lines/mm and thus the peak of diffraction efficiency is
around 1.3 $\mu$m at the Bragg condition when the incident angle of an
input beam to the normal of the grating surface is 15$^{\circ}$. The
measurements are performed at wavelengths from 0.9 $\mu$m to 1.6 $\mu$m.
The target temperature, size and line density of the grating, and
wavelengths investigated are nearly the same as those adopted for the
Fibre-Multi Object Spectrograph (FMOS; e.g., Kimura et al. 2003), which
is one of the next generation instruments for the 8.2m Subaru telescope
with commissioning expected in 2004: this instrument will be observing
at wavelengths from 0.9 $\mu$m to 1.8 $\mu$m, and a VPH grating will be
used as an anti-dispersing element in the near-infrared spectrograph
which is operated at $\sim$ 200 $K$ to reduce thermal noise.

\section{Pre-test at Room Temperature}

Before starting measurements using the cryogenic test facility (see next
section), we investigated the diffraction efficiency of the VPH grating
at a room temperature. The measurements were performed manually on the
optical bench. The procedures and results obtained are summarised below.

\subsection{Measurements}

\begin{figure}
\begin{center}
 \includegraphics[height=10cm,angle=-90,keepaspectratio]{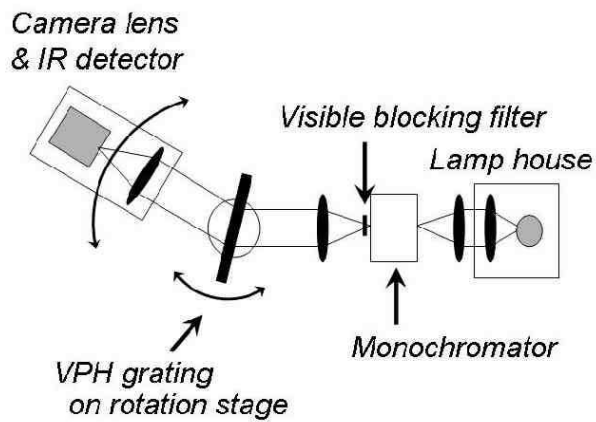}
 \caption{Schematic view of the overall configuration of the optics for
 the pre-test at a room temperature.}  \label{config}
\end{center}
\end{figure}

In Figure \ref{config}, the overall configuration of the optical
components used for the measurements is indicated (the detailed
information for the main components are listed in Table \ref{comps}).
Light exiting from the monochromator is collimated and used as an input
beam to illuminate the central portion of the VPH grating. The spectral
band-width of this input beam is set by adjusting the width of the
output slit of the monochromator. The slit width and the corresponding
spectral band-width were set to 0.5 mm and $\sim$ 0.01 $\mu$m,
respectively, throughout the measurements; the beam diameter was set to
$\sim$ 2 cm by using an iris at the exit of the lamp house.

The input beam is diffracted by the grating and the camera, composed of
lenses and a near-infrared detector (320 $\times$ 256 InGaAs array), is
scanned so as to capture the diffracted beam. The output slit of the
monochromator is thus re-imaged on the detector. Since the detector has
some sensitivity at visible wavelengths, a visible blocking filter which
is transparent at wavelengths longer than 0.75 $\mu$m is inserted after
the monochromator to reduce contamination of visible light from a higher
order.

The basic measurement procedures are as follows. First, the brightness
of the lamp and the wavelength of light exiting from the monochromator
are fixed (the brightness of the lamp is kept constant by a stabilised
power supply during the measurement cycle at a given wavelength), and
the total intensity included in the image of the slit is measured
without the VPH grating. Then, the VPH grating is inserted at an angle
to the optical axis, and the intensities of the zero and first order
($+1$) diffracted light are measured. The diffraction angle is also
recorded. Next, the grating is set at a different incident angle and
the intensities of the diffracted light and diffraction angles are
measured. After these measurements are repeated for all the incident
angles of interest, a different wavelength is chosen and the same
sequence is repeated.
The brightness of the lamp can be changed when moving from one
wavelength to another: a higher brightness was used at shorter
wavelengths because the system throughput is lower.

\subsection{Results}

\begin{figure}[h]
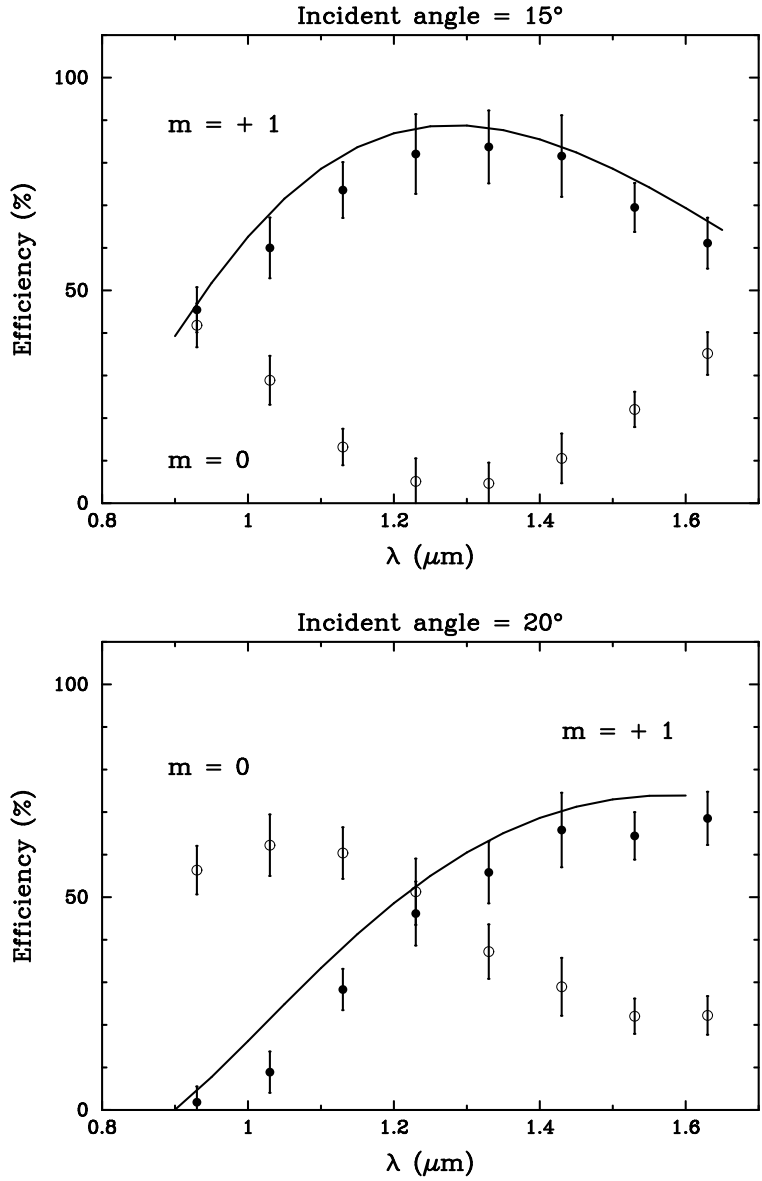

\begin{center}
 \includegraphics[height=10cm,angle=-90,keepaspectratio]{ntfig3.ps}

 \vspace{5mm}

 \includegraphics[height=10cm,angle=-90,keepaspectratio]{ntfig4.ps}

 \caption{Diffraction efficiency measured at a room temperature. The
 upper panel shows the efficiencies for the incident angle of
 15$^{\circ}$ and the lower panel is for the incident angle of
 20$^{\circ}$. Open and solid dots show the efficiency profiles for the 
 zero and first order ($+1$) diffracted light, respectively.
 theoretical prediction based on the coupled wave analysis. See text for
 details.}  \label{eff_c}
\end{center}
\end{figure}

In Figure \ref{eff_c}, the diffraction efficiencies measured are plotted
against wavelengths for the cases where incident angles are 15$^{\circ}$
(upper panel) and 20$^{\circ}$ (lower panel). Open and solid dots show
the efficiency profiles for the zero and first order ($+1$) diffracted
light, respectively. Since random errors are dominated by fluctuations
of the bias level of the detector on a short time scale ($\sim 0.1 - 1$
sec), the error bars plotted are calculated from a typical value of the
fluctuations. It is found from this figure that particularly for the
incident angle of 15 $^{\circ}$, the peak of the diffraction efficiency
reaches $\sim$ 80\% and the efficiency exceeds 50\% over the wavelength
range from 0.9 $\mu$m to 1.6 $\mu$m. The profiles can be well reproduced
by theoretical calculations based on the coupled wave analysis (Kogelnik
1969), which are shown by the solid lines in the figure. In these
calculations, a thickness of the dichromated gelatin layer of 12 $\mu$m
and a refractive index modulation amplitude of 0.05 are assumed. Energy
losses by surface reflections at boundaries between glass and air
($\sim$ 10\%) are also included.
Although the energy losses at boundaries between glass and gelatin are
likely to be much smaller because their refractive indices are very
similar, they might explain that the measured diffraction efficiency
tends to be slightly lower than the theoretical calculation.
(The energy lost by internal absorption of the dichromated gelatin layer
is estimated to be $\leq$ 1\% below 1.8 $\mu$m; e.g., Barden et al.
1998).

\section{Cryogenic Test}

In the following, we describe the measurements at 200 $K$ as well as
those at a room temperature, both of which were performed using the
cryogenic test facility as shown below. Throughout these measurements,
we used a different VPH grating from that used in the pre-test at room
temperature, although both have the same specifications.

\subsection{Measurements}

\begin{figure}
\begin{center}
 \includegraphics[height=10cm,angle=-90,keepaspectratio]{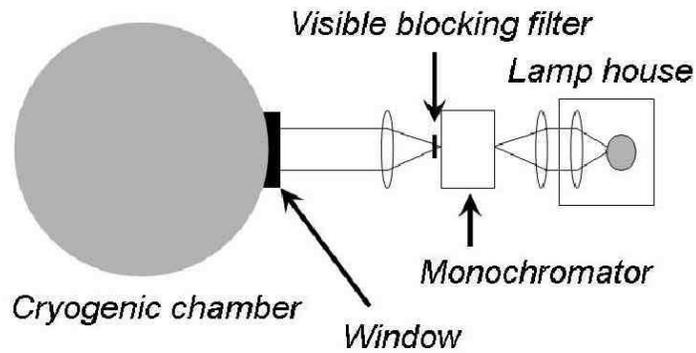}

 \vspace{5mm}

 \includegraphics[height=10cm,angle=-90,keepaspectratio,clip]{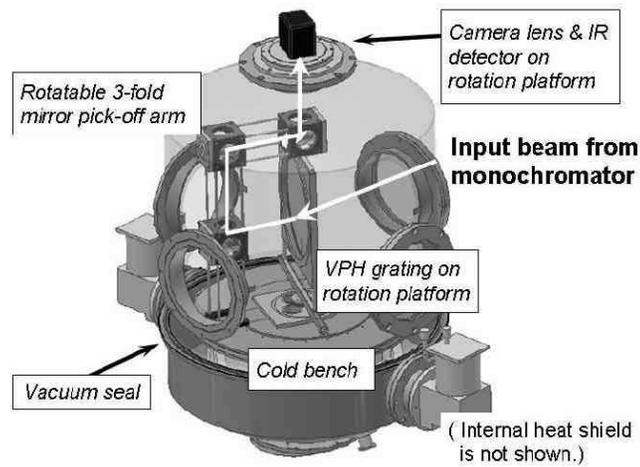}
 \caption{The upper panel shows schematic view of the overall
 configuration of the optics in the test setup. The lower panel shows
 schematic view of the components and the light path inside the
 cryogenic chamber. The internal heat shield is omitted for clarity.}
 \label{config1}
\end{center}
\end{figure}

\begin{figure}
\begin{center}
 \includegraphics[height=10cm,angle=-90,keepaspectratio]{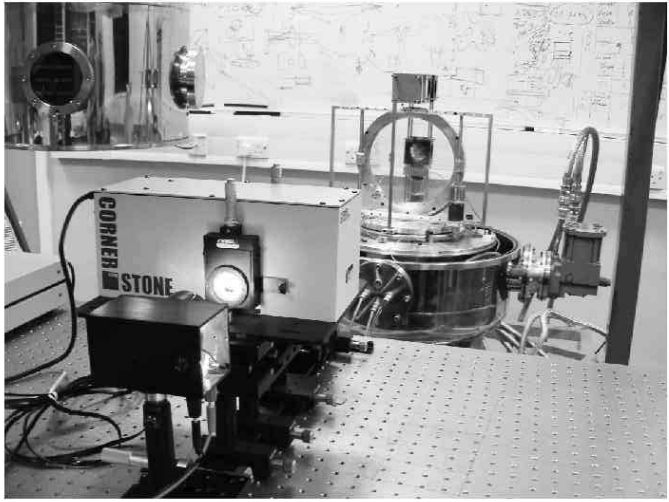}

 \includegraphics[width=10cm,keepaspectratio,clip]{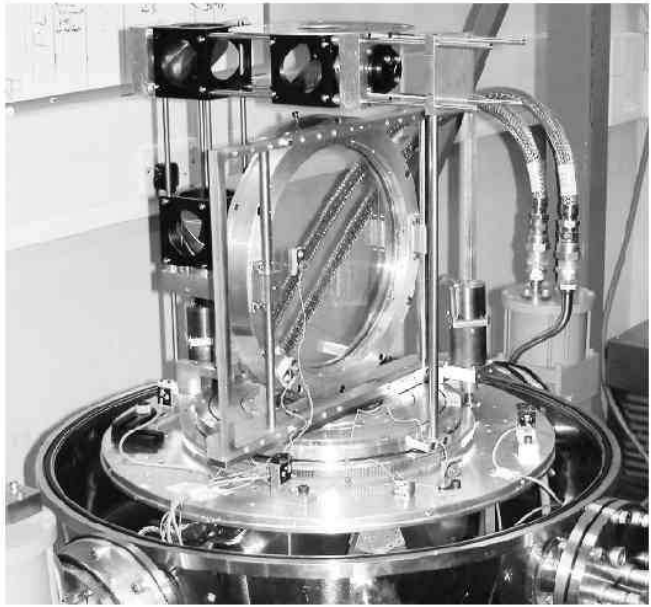}
 \caption{In the upper panel, picture of the facilities from the
 fore-optics side is shown. The vacuum vessel lid and heat shield for
 the cryogenic chamber have been taken off. In the lower panel, picture
 of the components inside the cryogenic chamber is shown.} \label{pic}
\end{center}
\end{figure}

In Figures \ref{config1}, schematic views of the fore-optics and the
optics inside the cryogenic chamber are indicated. Pictures of these
facilities are shown in Figure \ref{pic}. The fore-optics and light path
before the window of the cryogenic chamber and the camera are the same
as those used in the warm pre-test. The slit width and the spectral
band-width were set to 0.1 mm and $\sim$ 2.0 $\times 10^{-3}~\mu$m,
respectively, throughout the measurements, and the beam diameter was set
to $\sim$ 2 cm by using an iris before the window. The light path in the
cryogenic chamber is described as follows: the input beam illuminates
the central portion of the VPH grating and is diffracted by the
grating. The diffracted light is captured by scanning the pick-off arm
and is delivered to the camera on the top of the chamber by 3 pick-off
mirrors. The output slit of the monochromator is thus re-imaged onto the
detector.
This procedure enables measurements to be made at a variety of incident
and diffraction angles without having to mount the detector inside the
cryostat.

The measurement procedures are the same as those in the pre-test, except
that all the measurements were performed with the VPH grating in place.
We initially perform the measurements at a room temperature ($\sim$ 280
$K$). Then, we repeat the measurements at 200 $K$ before returning to
280 $K$ to repeat the cycle. When we cool the VPH grating, we monitor
the temperature of the grating with a sensor on the surface, close to
the edge of the grating but unilluminated by the input beam. When the
temperature reaches $\sim 200~K$, we switch off the compressor and cold
heads before starting the measurements. Although we do not have any
thermostatic systems to maintain a given temperature, it takes several
hours for the temperature of the grating to start increasing and go
above 200 $K$ after the compressor and cold heads are switched off. Thus
the temperature of the grating stays approximately at 200 $\pm$ 5 $K$
for the duration of the measurement cycle.

\subsection{Results and Discussions}

The following results were obtained when the incident angle was set to
$15^{\circ}$, which gives the peak of diffraction efficiency around 1.3
$\mu$m when satisfying the Bragg condition. We note that the same trends
are obtained from measurements when different incident angles are
adopted.

\subsubsection{Variation of diffraction efficiency}

\begin{figure}
\begin{center}
 \includegraphics[width=8cm,keepaspectratio]{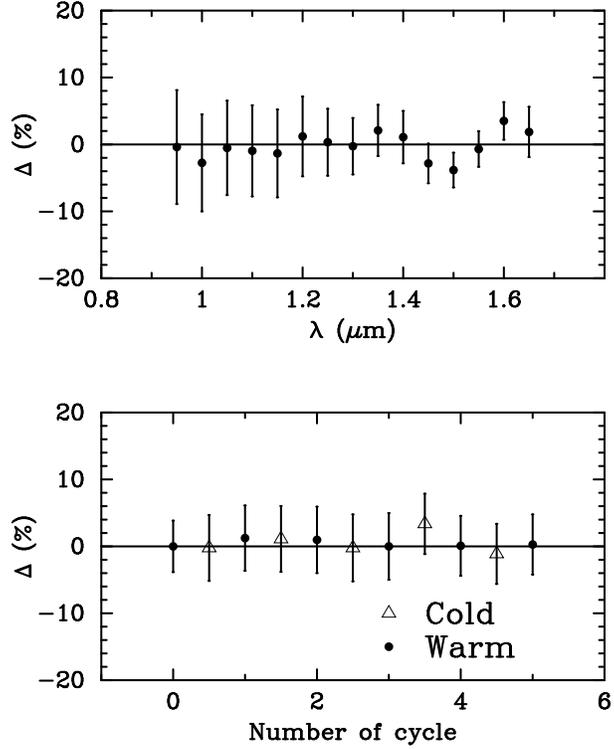}
 \caption{Comparison of diffraction efficiency at 200 $K$ with that at
 280 $K$. In the upper panel, differences of diffraction efficiencies in
 a sequence of measurements at 200 $K$ from those obtained in the first
 warm test (280 $K$) are plotted against wavelengths. The error bars are
 calculated from the typical fluctuation of the bias level of the
 detector. In the lower panels, the differences in diffraction
 efficiency as shown above are averaged over the wavelength range, and
 the averaged difference from the first warm test is plotted against
 cycle number. Open triangles and solid dots represent the data at 200
 $K$ and those at 280 $K$, respectively. The error bars indicate the
 standard deviation of a distribution of the differences around the
 average value.} \label{eff}
\end{center}
\end{figure}

In the upper panel of Figure \ref{eff}, differences of diffraction
efficiencies in a sequence of measurements at 200 $K$ from those
obtained in the first warm test (280 $K$) are plotted against
wavelengths. The error bars are calculated from the typical fluctuation
of the bias level of the detector. The errors are larger at shorter
wavelengths because the system throughput is lower so that
the bias level fluctuation is larger compared to the intensity of the
slit images (the brightness of the lamp is kept constant with a
stabilised power supply throughout the measurements at all the
wavelengths). It is found that the differences are close to zero at all
the wavelengths, suggesting that there is no significant variation in
profile of diffraction efficiency such as a global decrease in the
efficiency or a lateral shift of the peak.
In the lower panels, the differences in diffraction efficiency are
averaged over the wavelength range, and the averaged difference from the
first warm test is plotted against cycle number. Open triangles and
solid dots represent the measurements at 200 $K$ and those at 280 $K$,
respectively. The error bars indicate the standard deviation of a
distribution of the differences around the average value.
These results suggest that the diffraction efficiency of a VPH grating
is nearly independent of temperature, at least between 200 $K$ and 280
$K$, and that no significant deterioration is caused by a small number
of heating and cooling cycles.

\subsubsection{Variation of angular dispersion}

\begin{figure}
\begin{center}
 \includegraphics[width=8cm,keepaspectratio]{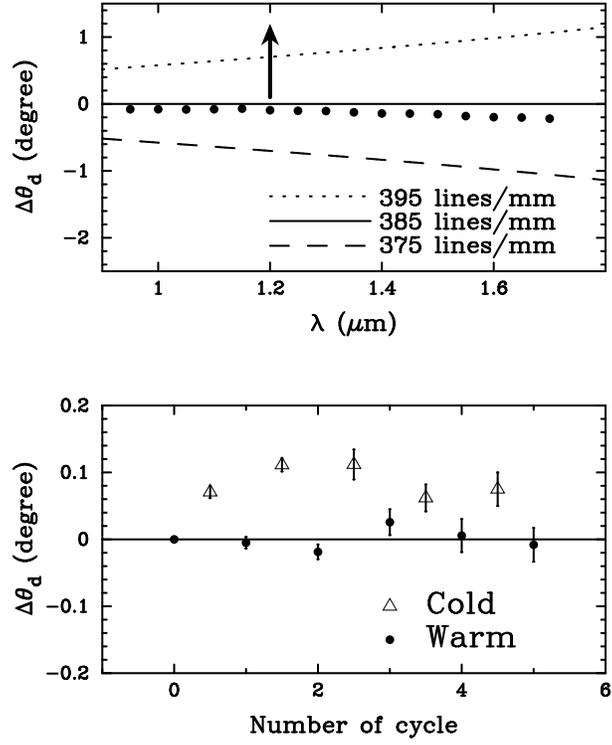}
 \caption{Relationship between diffraction angle and wavelength. In the
 upper panel, the differences from the prediction for 385 lines/mm are
 presented. If the gelatin layer shrinks with decreasing temperature,
 the line density would increase and the data points would shift upwards
 on this plot as the arrow shows. In the lower panel, comparison of
 diffraction angle at 200 $K$ with that at 280 $K$ is shown. See text
 for details. Note that the scale of the vertical axis in the lower
 panel is extended for clearer presentation.} \label{dispersion}
\end{center}
\end{figure}

In the upper panel of Figure \ref{dispersion}, difference of diffraction
angle from the prediction for the line density of 385 lines/mm (the
nominal line density of the VPH grating) is plotted against wavelength;
solid line (zero level) corresponds to the relationship between
diffraction angle and wavelength for the line density of 385 lines/mm.
Dotted and dashed lines indicate the relationships predicted for a line
density of 375 and 395 lines/mm, respectively. The data points show the
actual measurements at 280 $K$.\footnote{Although this plot may suggest
that the line density of the VPH grating is slightly smaller than 385
lines/mm, precise definition of the line density through measurements is
beyond the scope of this paper.} As the arrow shows, if the gelatin
layer shrinks with decreasing temperature, the line density would
increase and the data points would shift upwards on this plot.

In the lower panel, the difference of diffraction angle from that
measured in the first warm test is plotted against the number of
cycles. The symbols have the same meanings as those in Figure
\ref{eff}. Again, each data point represents the difference averaged
over the wavelength range, and the error bars indicate the standard
deviation of the distribution of the differences around the average
value. The lower panel suggests that at 200 $K$, the diffraction angle
is slightly larger than that at 280 $K$. This is equivalent to a slight
increase of the line density of the grating, and the simplest
explanation for this is a shrinkage of the grating with decreasing
temperature. By using this increment of diffraction angle ($\sim$
0.1$^{\circ}$), the amount of shrinkage is estimated to be $\sim 0.5$\%
of the diameter of the grating, which is consistent with the amount of
shrinkage of the glass substrate expected when the temperature is
decreased by 80 $K$ (the amount of shrinkage of gelatin would be larger
by an order of magnitude).
One needs to keep in mind, however, that only a small portion of the VPH
grating was illuminated throughout the measurements (the beam diameter
was $\sim$ 2 cm while the diameter of the VPH grating is $\sim$ 25 cm)
and thus a variation of the line density might be difficult to be
detected. Investigating larger portions over the VPH grating would be an
important future work.

\subsection{Comparison with other results}

Other cryogenic tests of VPH gratings are also in progress by the gOlem
group at Brera Astronomical Observatory (Bianco et al. 2003). Their
preliminary results suggest that diffraction efficiency is significantly
reduced ($\sim$ 20\% around the peak) at $\sim$ 200 $K$ compared with
room temperature,
which is inconsistent with our results.
One consideration here is that there is a significant difference in
speed of the cooling and heating processes between the tests of the
gOlem group and ours. In the gOlem case, they require only 1 hour to
cool a VPH grating down to their target temperature from a room
temperature (Zerbi 2001).
With the cryogenic chamber used for our experiments, it takes about 15
hours to cool down to 200 $K$ from 280 $K$. This may imply that rapid
cooling and/or heating can cause some deterioration of a VPH
grating. Further experiments are required in this area.

\section{Summary \& Conclusion}

In this paper, results from the cryogenic tests of a VPH grating at 200
$K$ are presented. The aims of these tests were to see whether
diffraction efficiency and angular dispersion of a VPH grating are
significantly different at a low temperature from those at a room
temperature, and to see how many cooling and heating cycles the grating
can withstand. We have completed 5 cycles between room temperature and
200 $K$, and find that diffraction efficiency and angular dispersion are
nearly independent of temperature. This result indicates that VPH
gratings can be used in spectrographs cooled down to 200 $K$ such as
FMOS without any significant deterioration of the performance.

In future, we will be trying more cycles between 200 $K$ and 280 $K$ to
mimic more realistic situations of astronomical use. Measurements at a
much lower temperature (e.g., $\sim$ 80 $K$) will also be necessary to
see whether VPH gratings are applicable to spectrographs for use in the
$K$-band. We will report on these issues in a forthcoming paper.

\acknowledgements

We thank colleagues in Durham for their assistance with this work,
particularly Paul Clark, John Bate, and the members of the mechanical
workshop. We are also grateful to the anonymous referee for the 
comments to improve our paper. This work was funded by PPARC Rolling
Grant (PPA/G/O/2000/00485).

\clearpage

\begin{table}
\begin{center}
\caption{The main components used for the measurements.}\label{comps}
\begin{tabular}{llll} \hline
 & Manufacturer & Product ID & \multicolumn{1}{c}{Comments} \\ \hline
Light source            & Comar             & 12 LU 100       & 
Tungsten-halogen lamp \\
Monochromator           & Oriel Instruments & Cornerstone 130 & 
600 lines/mm grating, Blaze at 1 $\mu$m \\
                        &                   & Model 74000     & \\
Visible blocking filter & Comar             & 715 GY 50       &
 Transparent at $\lambda \geq 715$ nm \\
Near-infrared detector  & Indigo Systems    & Alpha$-$NIR     &
 320 $\times$ 256 InGaAs array \\ \hline
\end{tabular}
\end{center}
\end{table}

\clearpage

\begin{thebibliography}{}
\bibitem[1998]{barden1}
 Barden, S. C., Arns, J. A., \& Colburn, W. S. 1998, Proc. SPIE, 3355,
              866

\bibitem[2000]{barden2}
 Barden, S. C., Arns, J. A., Colburn, W. S., \& Williams, J. B. 2000,
              PASP, 112, 809

\bibitem[2003]{bianco}
 Bianco, A., Molinari, E., Conconi, P., et al. 2003, Proc. SPIE,
	      Vol. 4842, 22

\bibitem[2002]{bridges}
 Bridges, T., Aa$\Omega$ Team. 2002, Anglo-Australian Observatory Epping
              Newsletter, 100, 20

\bibitem[2003]{kimura}
 Kimura, M., Maihara, T., Ohta, K., Iwamuro, F., Eto, S., Iino, M.,
 Mochida, D., Shima, T., Karoji, H., Noumaru, J., Akiyama, M., Brzeski,
 J., Gillingham, P. R., Moore, A. M., Smith, G., Dalton, G. B., Tosh,
 I. A. J., Murray, G. J., Robertson, D. J., \& Tamura, N. 2003,
 Proc. SPIE. 4841, 974

\bibitem[1969]{kogelnik}
 Kogelnik, H. 1969, Bell System Tech. J., 48, 2909

\bibitem[2001]{zerbi}
 Zerbi, F. M. 2001, ``Preliminary tests on VPH in Cryogenic environment''
on the web page: http://golem.merate.mi.astro.it/projects/vph/cryo/cryo.html

\end{thebibliography}
\end{document}